\documentclass{PoS}

\title{Quasi-particle perspective on critical end-point}

\ShortTitle{Quasi-particle perspective on critical end-point}

\author{Marcus Bluhm\\
  Institut f\"ur Strahlenphysik, Forschungszentrum Dresden-Rossendorf, \\ 
  PF 510119, 01314 Dresden, Germany\\
  E-mail: \email{m.bluhm@fzd.de}}
\author{\speaker{Burkhard K\"ampfer}\\
        Institut f\"ur Strahlenphysik, Forschungszentrum Dresden-Rossendorf, \\ 
        PF 510119, 01314 Dresden, Germany\\
        Institut f\"ur Theoretische Physik, Technische Universit\"at Dresden, 01062 Dresden, Germany\\
        E-mail: \email{kaempfer@fzd.de}}

\abstract{Within a quasi-particle model for the equation of state of strongly interacting matter 
for two quark flavors 
we include phenomenologically features of the QCD critical point and discuss its impact on the equation 
of state. In particular, we investigate the influence on the quark number susceptibility and the pattern of 
isentropic trajectories which describe the evolutionary paths of matter during the hydrodynamical expansion of 
a heavy-ion collision.}

\FullConference{Critical Point and Onset of Deconfinement\\
                 July 3-6 2006\\
                 Florence, Italy}

\begin{document}

\section{Introduction}

The phase diagram of strongly interacting matter shows an astonishing richness (cf. \cite{Raja,Risch} for some excellent 
overviews). At finite temperature 
$T^{CP}\sim \mathcal{O}(160\, \rm{MeV})$ and non-zero baryo-chemical potential 
$\mu_B^{CP}\sim \mathcal{O}(360\, \rm{MeV})$ \cite{Fodor,Gavai} a first-order phase transition line (for $\mu_B > \mu_B^{CP}$) 
separating confined from deconfined matter terminates. 
While the exact location $(T^{CP},\mu_B^{CP})$ of this end point, the critical point (CP) being of second order, 
is matter of current investigations (see comparison in \cite{Stephanov}, showing in fact a strong quark mass $m_q$ 
dependence \cite{Forc,Schm}), its existence for finite quark masses seems 
to be guaranteed by general arguments \cite{Stephanov,Halasz} relying 
on universality arguments. Note, however, the recent discussion \cite{deForc,Phil}, where an alternative scenario 
without CP is discussed. We take here the attitude to assume, in line with \cite{Stephanov}, the existence of a CP and 
discuss phenomenologically some of its possible implications. Besides, 
the extension of the critical region, which might be crucial for searches of CP 
in future heavy-ion experiments, is fairly unknown \cite{Scha,Hatta}. 
For massive quarks, the CP belongs to the universality class of the 3-dimensional 
Ising model which was numerically verified by first-principle QCD simulations (dubbed lattice QCD) for the volume dependence 
of the Binder cumulant \cite{Forc,Kar}. 
In the (unphysical) limit of vanishing quark masses, the CP turns into a tricritical point belonging 
to the universality class of the Heisenberg $\mathcal{O}(4)$ model in three dimensions \cite{Pis} while the 
transition line continues as second-order phase transition line of critical points up to $\mu_B=0$. 

For finite $m_q$ and $\mu_B < \mu_B^{CP}$, the sequence of first-order phase transitions turns into a cross over. Still, in this 
region a rapid increase of the entropy density $s$ or energy density $e$ in a narrow temperature interval 
or a peak in a suitable susceptibility allows 
for the determination of a pseudo-critical line defining the deconfinement temperature $T_c$ at $\mu_B=0$. 
At $T_c$, the chiral condensate $<\bar{q}q>$ as order parameter of chiral symmetry breaking and the expectation 
value of the Polyakov-loop as order parameter of deconfinement also rapidly change. 
For realistic quark masses reproducing the physical hadron spectrum, the transition of confined (hadron) matter 
to deconfined (quark-gluon) matter seems to be indeed a cross over \cite{Fodor,Laer}. This is in contrast to pure gauge 
theory, which displays at $T_c$ a first-order phase transition \cite{Risch,Karschi}. The quark mass dependence of the 
transition characteristics at $T_c$ is discussed in some detail in \cite{Risch,Laer}. 

The CP as fundamental issue of QCD is matter of current investigations both theoretically and experimentally. 
In particular, observable consequences are discussed \cite{Stephanov,Asa}. Concerning the hydrodynamical 
description of the expansion stage in heavy-ion collisions, one might ask to what extent the equation of state (EoS) 
becomes modified by CP. 

Our paper is organized as follows. In section 2, we briefly review our quasi-particle model for the EoS 
of strongly interacting matter and compare with recent lattice QCD simulations for two quark flavors, in particular 
the isentropic evolutionary paths during the hydrodynamical expansion in heavy-ion collisions. In section 3, we discuss 
the influence of including critical point effects phenomenologically into our model on the EoS. Our results are 
summarized in section 4. 

\section{Quasi-Particle Model and EoS for ${\bf N_f=2}$}

The pressure $p$ of $N_f=2$ light quark flavors in our quasi-particle model (QPM) as function of temperature $T$ and one chemical 
potential $\mu_q=\mu_B/3$ reads $p(T,\mu_q)=\sum_{a=q,g}p_a - B(\Pi_{q,g}(T,\mu_q))$ (cf. \cite{Peshier} for details) 
with partial pressures of quarks, $p_q$, and transverse gluons, $p_g$, predominantly propagating on-shell with dispersion 
relation $\omega_a=\sqrt{k^2+\Pi_a}$. This quasi-particle picture is motivated by a chain of 
feasible approximations starting 
from the Luttinger-Ward approach to QCD (cf. \cite{Bluhm1,Bluhm2} for details). $\Pi_a$ are given by the asymptotic 
expressions of the gauge-independent hard-thermal loop / hard-dense loop self-energies \cite{leBellac} and $B$ is 
determined by thermodynamic self-consistency and stationarity of the thermodynamic potential under functional variation 
with respect to the self-energies, $\delta p/\delta \Pi_a = 0$ \cite{Gorenstein}. Non-perturbative effects are 
accommodated in the model by replacing the running coupling $g^2$ in $\Pi_a$ by an effective coupling $G^2$ which needs 
to be determined for arbitrary $T$ and $\mu_q$. This can be achieved by solving a flow equation following from 
Maxwell's relation \cite{Peshier,Bluhm2} as Cauchy-problem together with a convenient parametrization of $G^2$ at 
$\mu_q=0$ \cite{Bluhm3}. 

From the pressure $p$, other thermodynamic quantities such as entropy density $s$ or baryon density $n_B$ follow 
straightforwardly. Assuming local entropy and baryon number conservation during the hydrodynamical expansion in 
heavy-ion collisions, the evolutionary paths of individual fluid elements are 
represented by isentropic trajectories $s/n_B = const$ in 
the $T$ - $\mu_B$ plane. Recently $n_B$, $s$ or energy density $e$ have been evaluated in first-principle numerical 
simulations of QCD \cite{All05,Ejiri} as Taylor series expansions 
\begin{eqnarray}
  \label{e:sn}
  s(T,\mu_B)  & = & T^3\sum_{n=0}^\infty s_n(T)\left(\frac{\mu_B}{3T}\right)^n \,,\\ 
  e(T,\mu_B)  & = & T^4\sum_{n=0}^\infty e_n(T)\left(\frac{\mu_B}{3T}\right)^n \,, 
\end{eqnarray}
including terms up to order $(\mu_B/T)^6$ (terms of order $(\mu_B/T)^8$ are currently calculated). 
Here, $e_n(T) = 3 c_n(T) + c_n'(T)$ and $s_n(T) = (4-n) c_n(T) + c_n'(T)$ where $c_n'(T)=T dc_n(T)/dT$ and 
\begin{equation}
  \label{e:coeff1}
  c_n(T) = \left.\frac{1}{n!} \frac{\partial^n (p/T^4)}{\partial  (\mu_q/T)^n}\right|_{\mu_q = 0} 
\end{equation}
are the expansion coefficients for the pressure $p$ \cite{All05} which can be derived from the QPM expression for $p$ via 
(\ref{e:coeff1}). Using these expansions up to order $(\mu_B/T)^6$, we calculate 
the isentropic trajectories for $s/n_B = 300$, $45$ and compare with lattice QCD results \cite{Ejiri} in Fig.~1. 
\begin{figure}[h]
  \centering{
  \includegraphics[scale=0.29,angle=-90.]{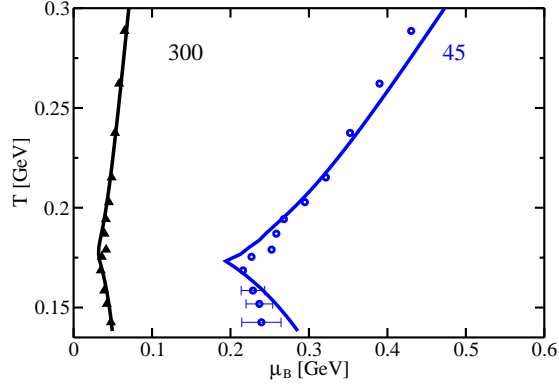}}
  \caption{Isentropic trajectories: Comparison of $N_f=2$ lattice QCD results \cite{Ejiri} for $s/n_B = $ 300 (triangles) 
    and 45 (circles) with the corresponding QPM results. In both cases, a truncated Taylor series expansion up to order 
    $(\mu_B/T)^6$ for $s$ and $n_B$ is applied. The deconfinement temperature at $\mu_B=0$ is 
    set to $T_c=175$ MeV \cite{Ejiri}.}
  \label{fig:isentrops}
\end{figure}
The agreement is fairly good (cf. \cite{Buda} for a discussion of these results). 
In \cite{Bluhm3}, we reported an impressive agreement between QPM and lattice 
QCD results for $c_n(T)$. As the Taylor expansion coefficients of $s$ and $e$ contain both, $c_n$ and $c_n'$, they serve for 
a more sensitive test of our model. We compare the QPM results for $s_{2,4}$ and $e_{2,4}$ with lattice QCD results 
\cite{Ejiri} in Fig.~2 finding a fairly good agreement. 
\begin{figure}[h] 
  \includegraphics[scale=0.28,angle=-90.]{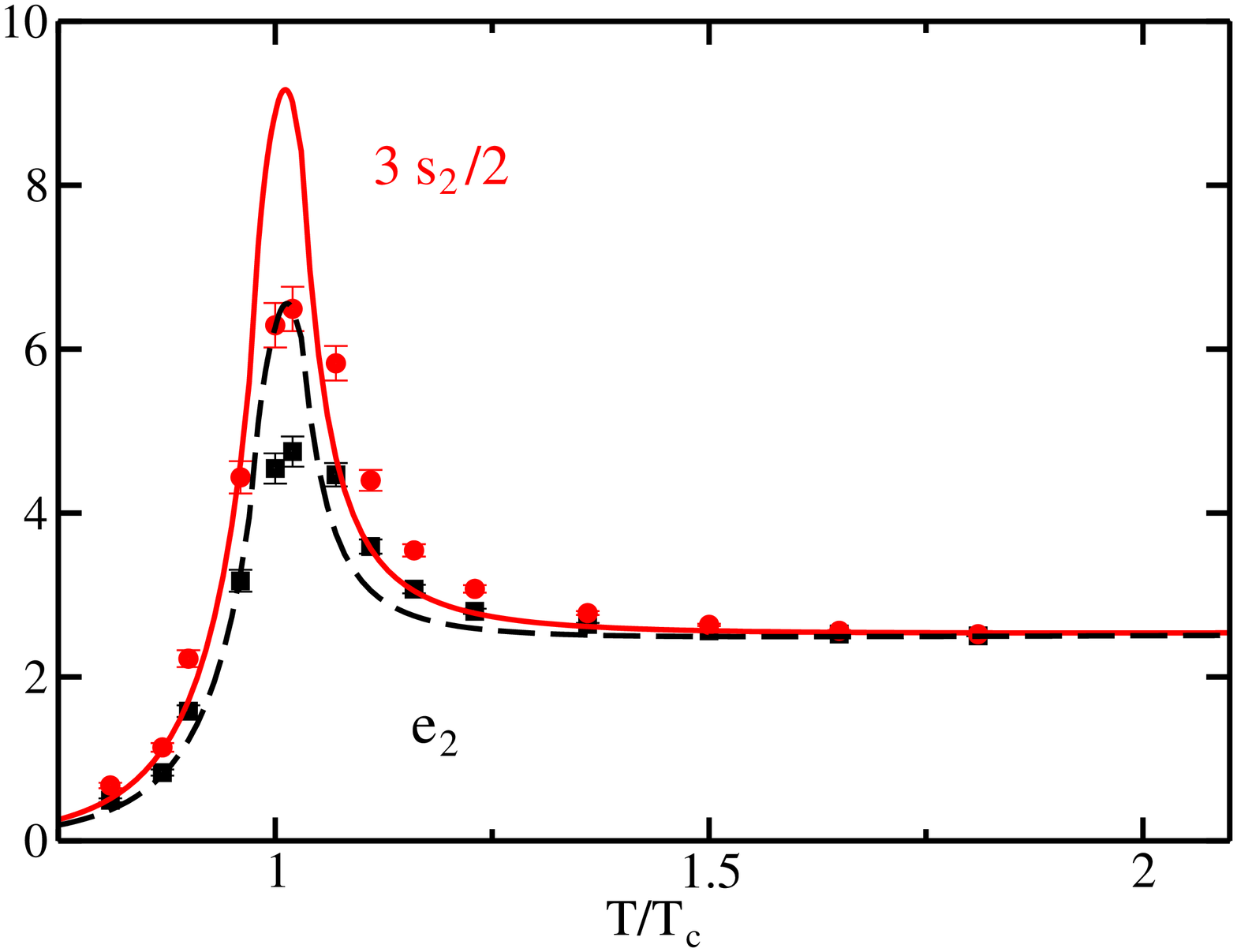}
  \includegraphics[scale=0.28,angle=-90.]{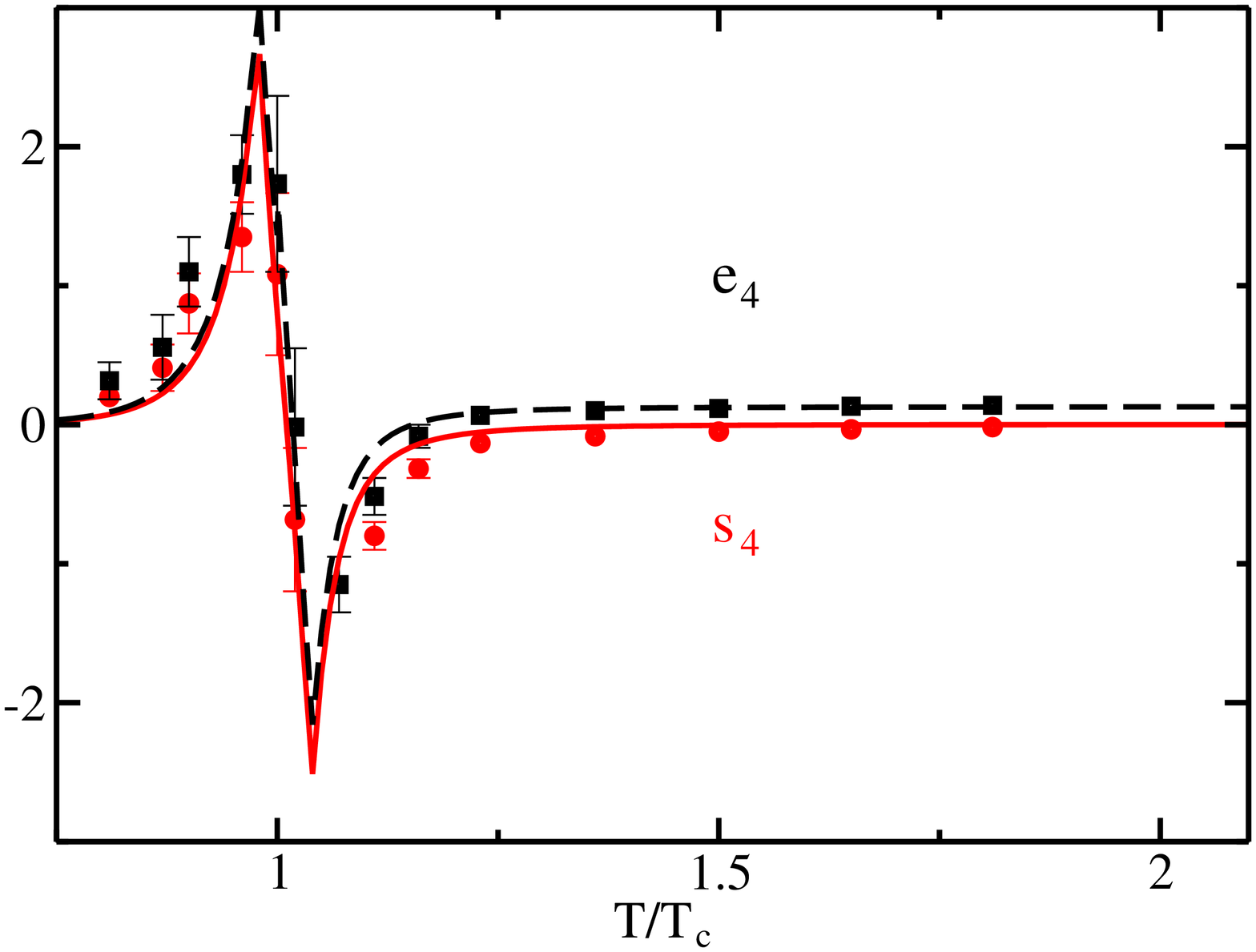}
  \caption{Comparison of our QPM with lattice QCD results \cite{Ejiri} for $e_n(T)$ (squares) and $s_n(T)$ (circles) 
    as function of $T/T_c$ for $N_f=2$; left (right) panel for $n=2\,(4)$. 
    QPM parameters chosen according to \cite{Bluhm3,Buda}. The pronounced structures in the vicinity of $T=T_c$ follow 
    from a curvature change in the parametrization of $G^2(T,\mu_q=0)$ (cf. \cite{Bluhm3}).} 
  \label{fig:coeffs2}
\end{figure}

It should be emphasized that the QPM does not include any critical behavior. However, the adjustment to lattice QCD 
data for $p(T,\mu_q=0)$ \cite{Kar1} forces us to build in a change of the slope of the effective coupling $G^2(T,\mu_q=0)$: 
Above $T_c$, the coupling resembles a regularized perturbative logarithmic behavior, while below $T_c$ a linear 
dependence on $T$ is required to describe the data. This change of the slope of $G^2(T,\mu_q=0)$ at $T_c$ generates 
the structures seen in Fig.~2, as higher order derivatives of $G^2$ enter. With the above reasoning we now attempt to 
supplement our QPM by structures causing a critical behavior. 

\section{Including the Critical Point \label{sec:CPInclusion}}

Following \cite{Nonaka05}, we aim at incorporating phenomenologically CP features into the EoS \linebreak parametrization 
and choose as starting point the decomposition of the entropy density $s$ 
into a regular part $s_{reg}$ and a singular part $s_{sing}$ which is related to phase transitions and critical 
phenomena  \cite{Gebhardt80}. For $s_{reg}$ we use our QPM entropy density, 
and $s_{sing}$ is constructed from the parametric representation \cite{Nonaka05,Guida97} of the Gibbs' 
free energy density $G(r,h)$ with critical behavior belonging to the universality class of the 3-dimensional Ising model. 
The order parameter of the 3-dimensional Ising model is the magnetization $M(r,h)$. $G(r,h)$ and $M(r,h)$ are 
functions of reduced temperature $r=(T-T_c)/T_c$ and external magnetic field $h$. From \cite{Nonaka05,Guida97}, the parametric 
form of $G$ in terms of new parameters $R$ and $\theta$ reads
\begin{equation}
  \label{equ:free}
  G = h_0 M_0 R^{2-\alpha} g(\theta) - Mh 
\end{equation}
with $M=M_0R^\beta \theta$ and $M_0$, $h_0$ are normalization constants. The variables of the 3-dimensional Ising model 
are defined by 
\begin{eqnarray}
  \label{equ:coo1}
  r & = & R \left(1-\theta^2\right) \,,\\
  \label{equ:coo2}
  h & = & h_0 R^{\beta\delta} \sum_{i=0}^2 a_{2i+1} \theta^{2i+1} \,,
\end{eqnarray}
where $R\ge 0$, $|\theta |\ge 1.154$, the critical exponents for the 3-dimensional Ising model 
read $\beta = 0.325$, $\delta = 4.80$, $\alpha = 2-\beta\left(1+\delta\right)$ and 
$a_1=1$, $a_3=-0.76201$, $a_5=0.00804$. The function $g(\theta)$ follows from solving the differential equation 
\begin{equation}
  \label{equ:diffequ}
  \sum_{i=0}^2 a_{2i+1} \theta^{2i+1} \left(1-\theta^2 + 2\beta\theta^2 \right) = 2 (2-\alpha) \theta g(\theta) + 
  (1-\theta^2) g'(\theta) 
\end{equation}
with integration constant $g(\theta =1)=0.04242$. Following these definitions, $M(r,h)$ shows the correct critical behavior 
$M(r=0,h)\sim {\rm sgn}(h) |h|^{1/\delta}$, $M(r<0,h=0^+)\sim |r|^\beta$ close to $r=0$, $h=0$ when choosing $M_0$, $h_0$ 
appropriately. 
\begin{figure}[ht]
  \begin{center}
  \includegraphics[scale=0.26,angle=-90.]{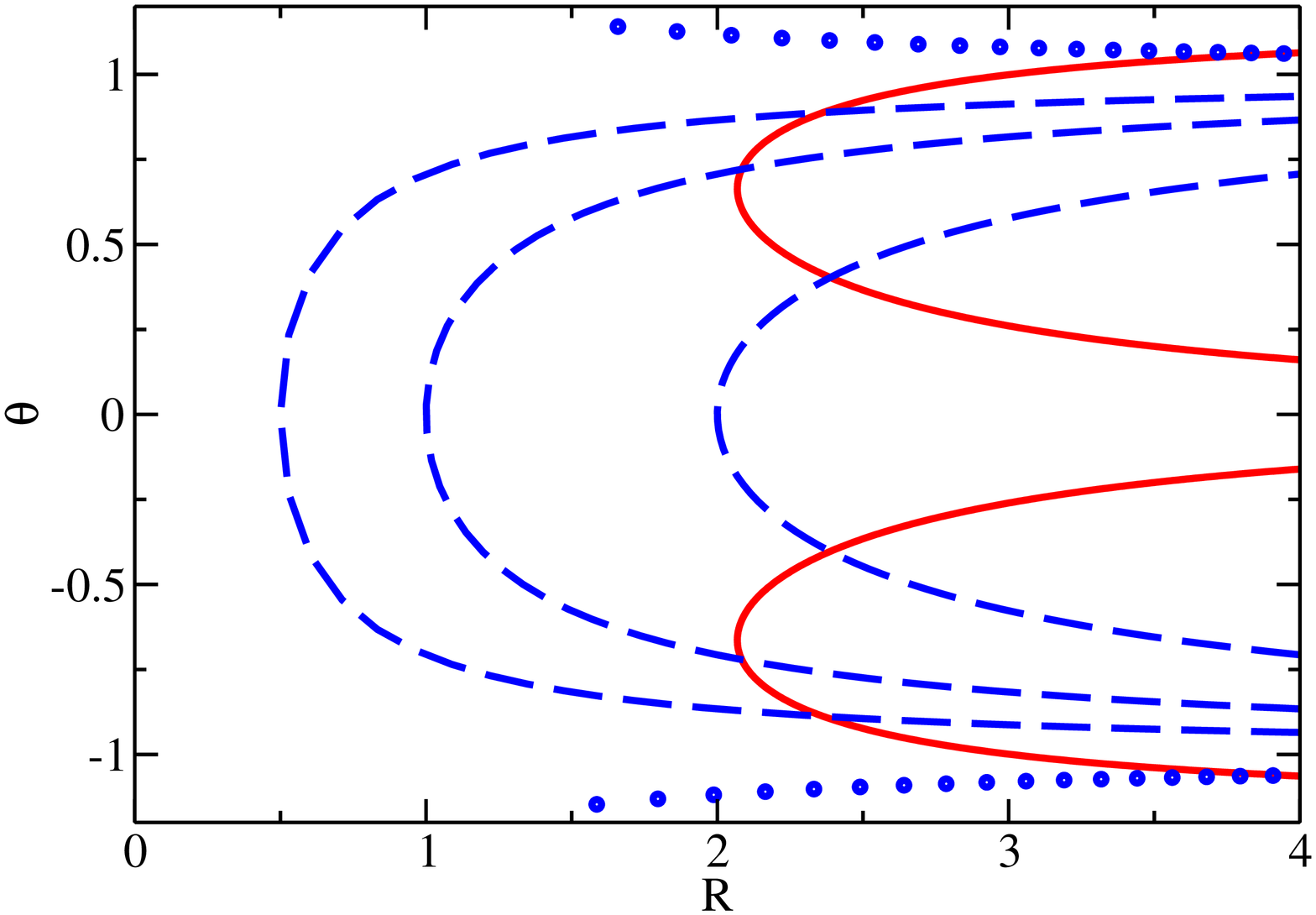}
  \includegraphics[scale=0.26,angle=-90.]{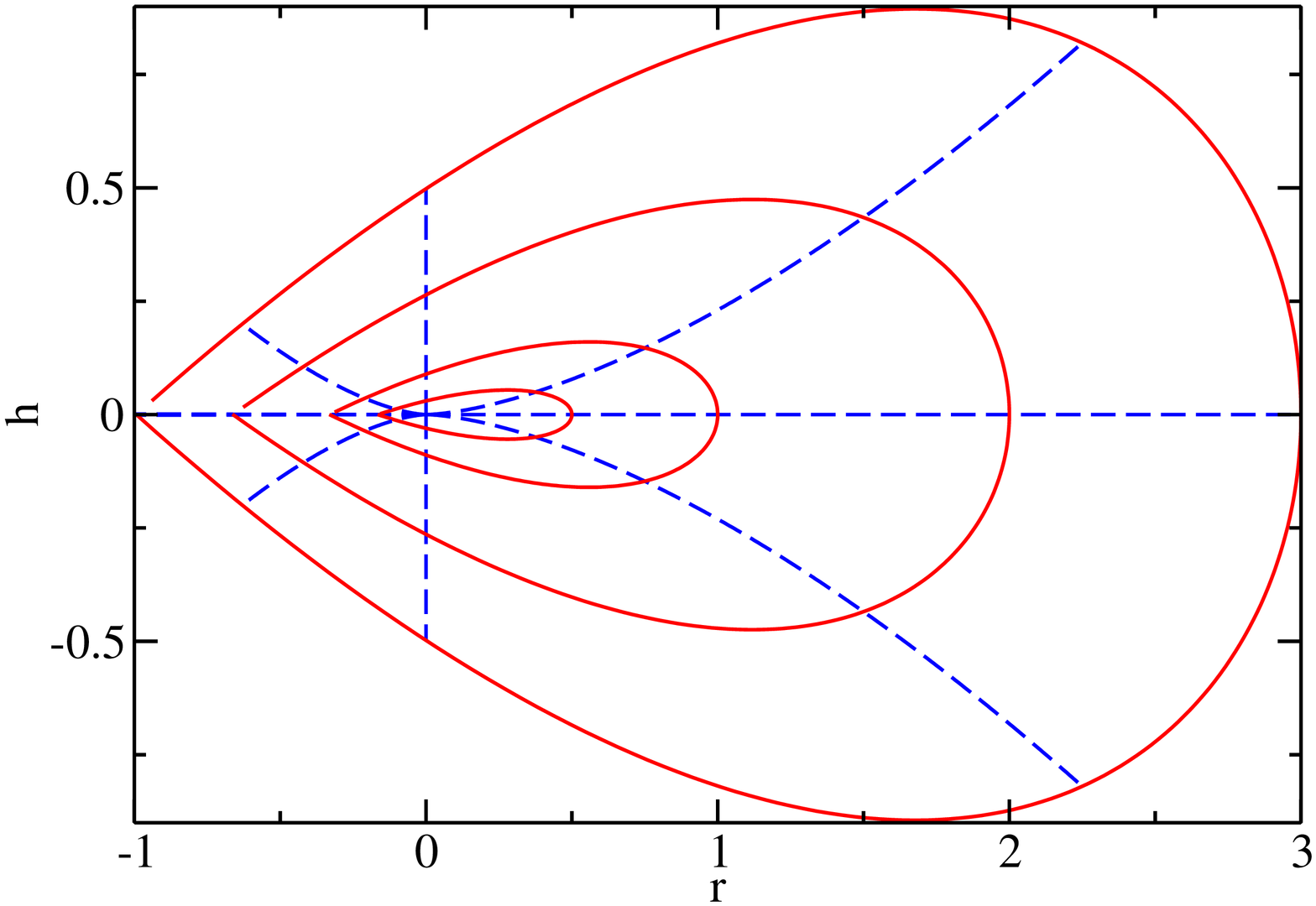}
  \caption{Mapping ($r,h$) $\leftrightarrow$ ($R,\theta$) of the parametric representation of $G$ in \cite{Nonaka05,Guida97} according to 
    (3.2, 3.3). Left panel: solid curves denote lines of constant $h=0.5$ ($-0.5$) for the upper (lower) curve. 
    Dashed curves depict lines of constant $r>0$ ($r=0.5, 1, 2$ from left to right), whereas dotted lines represent constant 
    negative $r=-0.5$. Right panel: solid curves denote lines of constant $R=0.5, 1, 2, 3$ from inner to outer ring, 
    dashed curves lines of constant $\theta$. For $h<0$, $\theta$ is negative ($\theta = -1.1, -1$ (here $r=0$) and $-0.5$ from 
    left to right); for $h>0$, $\theta$ is positive ($\theta = 1.1, 1$ (here $r=0$) and $0.5$ from left to right). For $h=0$, 
    $\theta = \pm 1.154$ for $r<0$ and $\theta = 0$ for $r>0$. By construction, the parametrization is continuous for $r>0$, whereas 
    the discontinuity of the parametrization in the region of $r<0$ generates a first-order phase transition for $\mu_B > \mu_B^{CP}$ 
    when choosing the coordinate system ($r,h$) in the $T$ - $\mu_B$ plane appropriately, cf. Fig. 4.} 
  \label{fig:CEPcoord2}
  \end{center}
\end{figure}
In Fig.~3, we illustrate the mapping of the variables of the 3-dimensional Ising model ($r,h$) on the parametric representation 
($R,\theta$) of the Gibbs' free energy density $G$ constructed in \cite{Nonaka05,Guida97}. 

CP is located at $r=0$ and $h=0$, cf. Fig.~4, and the coordinate system is orientated such that by construction 
the region of positive $r$ denotes the cross over region, whereas the region of $r<0$ defines the region of 
first-order phase transitions. 
\begin{figure}[h]
  \begin{center}
  \includegraphics[scale=0.29,angle=-90.]{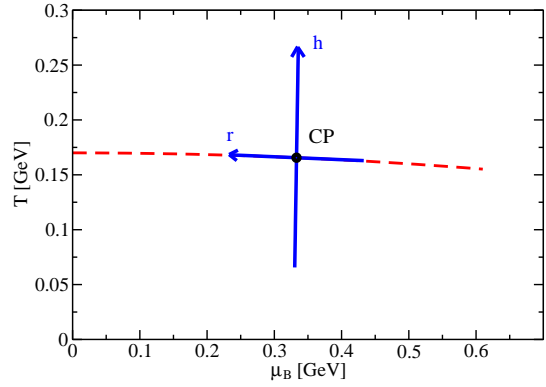}
  \caption{Visualization of the coordinate transformation between ($T,\mu_B$) and ($r,h$). CP is located according to 
    \cite{Fodor} on the estimated phase boundary $T_c(\mu_B)$. The coordinate system ($r,h$) is located such that 
    $r=0$, $h=0$ at CP. The $r$-axis is defined to be tangential to $T_c(\mu_B)$ at CP such that $r>0$ denotes the 
    cross over region, $r<0$ the region of first-order phase transitions. We put the $h$-axis perpendicular to the 
    $r$-axis. Note, that with increasing $\mu_B$, $r$-axis and phase boundary increasingly deviate from each other.}
  \label{fig:CEPcoord1}
  \end{center}
\end{figure}
The phase boundary is estimated by $T_c(\mu_B)=T_c\left(1+\frac{1}{2}d(\frac{\mu_B}{3T_c})^2\right)$ with $d=-0.122$ 
according to \cite{Forc,Alli} which agrees also fairly well with the curvature when solving for the flow equation 
of our QPM emanating at $T_c=175$ MeV. 

The singular part of the entropy density near the QCD critical point can be constructed from the dimensionless quantity 
\begin{equation}
  \label{equ:A1}
  S_c(T,\mu_B) = - \sqrt{\Delta T_c^2 + \Delta\mu_{B,c}^2}\left(\frac{\partial G}{\partial T}\right)_{\mu_B} .
\end{equation}
Here, $\Delta T_c$ and $\Delta\mu_{B,c}$ approximate the extension of the critical region in $T$ and $\mu_B$ directions, respectively, 
provided the CP is located at small $\mu_B$, by quantifying the rapidity with which $S_c(T,\mu_B)$ changes across the 
r-axis that mimics the phase transition line. From Fig.~4 it also becomes clear that this phenomenological construction of 
CP features into the model is limited in accuracy by the increasing discrepancy between the $r$-axis and the actual phase 
boundary between hadronic and quark-gluon matter as $\mu_B$ increases. Thus, this procedure works best if CP is located in 
the flat region of the phase border line at small $\mu_B$. 
The difficult task is to map the QCD variables $T$ and $\mu_B$ on $r$ and $h$ and to evaluate (\ref{equ:A1}) from 
(\ref{equ:free}). These relations were elaborated in the pioneering work \cite{Nonaka05} the interested reader is referred 
to for the explicit form. 

In order to illustrate CP effects on the EoS, in particular on the pattern of isentropic trajectories, we 
consider here $N_f=2$ and the CP located according to \cite{Fodor} at $\mu_B^{CP}=360$ MeV. Since we use 
$T_c(\mu_B=0)=175$ MeV (\cite{Ejiri}, see also \cite{Cheng}) 
we find by solving the flow equation \cite{QM05} a slightly different $T^{CP}=170$ MeV 
(compared to \cite{Fodor}). For the regular contribution $s_{reg}$ we employ 
(\ref{e:sn}) including $s_{0,2,4,6}(T)$ as calculated from our QPM with parameters as applied in \cite{Bluhm3,Buda} for 
describing $c_i(T)$ (cf. Fig.~2 for $s_{2,4}(T)$). 
When constructing $s_{sing}$ from the dimensionless quantity $S_c(T,\mu_B)$ in (\ref{equ:A1}), one important constraint 
to be satisfied is a vanishing net baryon density $n_B$ at $\mu_B=0$. From standard thermodynamic relations one finds 
\begin{equation}
  \label{equ:A2}
  n_B(T,\mu_B) - n_B(0,\mu_B) = 
  \int_0^T \frac{\partial s(T',\mu_B)}{\partial\mu_B} dT' .
\end{equation}
Assuming a finite extension of the critical region around CP, the integration constant $n_B(0,\mu_B)$ is completely 
determined by the regular part of the thermodynamic potential. Obeying the correct dimension of an entropy density, 
we therefore choose $s_{sing}$ proportional to the second-order Taylor expansion coefficient $c_2(T)$ from (\ref{e:coeff1}) 
(cf. \cite{Bluhm3}), 
\begin{equation}
  \label{equ:A3}
  s_{sing}(T,\mu_B) = \frac{2}{9} A c_2(T) \mu_B^2 T  tanh\,S_c(T,\mu_B) .
\end{equation}
Here, $A$ denotes the relative strength between regular and singular contributions and $tanh\,S_c$ binds $S_c(T,\mu_B)$ 
between $-1$ and $+1$ for convenience. 

In order to describe lattice QCD results, one has to adjust the remaining parameters $\Delta T_c, \,\Delta\mu_{B,c}$ and 
$A$ according to $s_{lQCD} = s_{reg} + s_{sing}$, where for small $\mu_B$ lattice QCD results have already been successfully 
described by the regular part alone. A possible solution is choosing the parameters in 
(\ref{equ:A1},~\ref{equ:A3}) as $\Delta T_c = \Delta\mu_{B,c} = 2$ MeV mimicking a small critical region and $A=0.1$. 
This solution implies that bulk thermodynamic quantities like $p(T,\mu_B)$, $e(T,\mu_B)$ or $s(T,\mu_B)$ 
are represented to a large extent by the regular contribution, i.~e. by the QPM parametrization, 
as evident in Fig.~5 (left panel) for the isentropic 
\begin{figure}[h]
  \begin{center}
  \includegraphics[scale=0.26,angle=-90.]{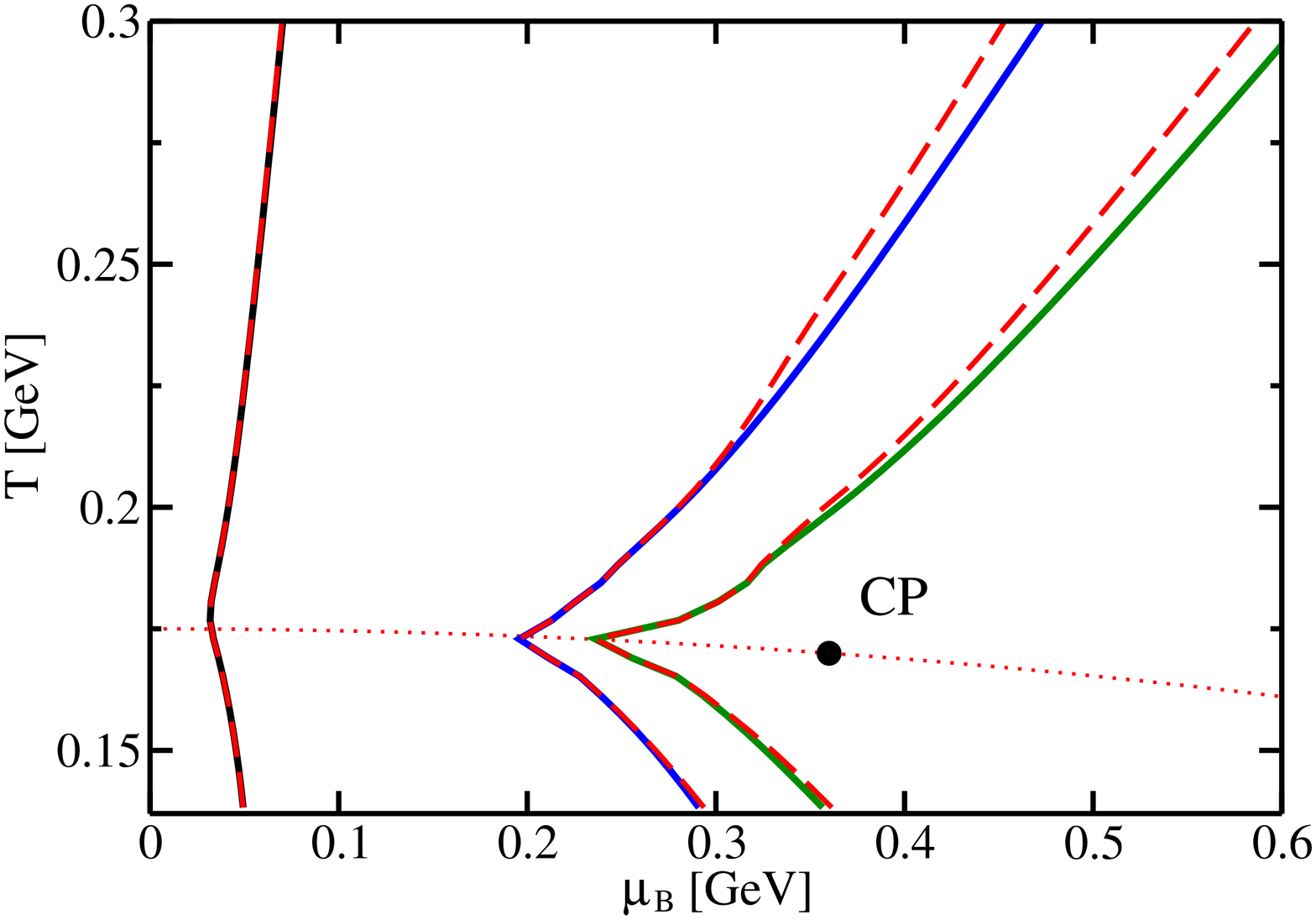}
  \includegraphics[scale=0.26,angle=-90.]{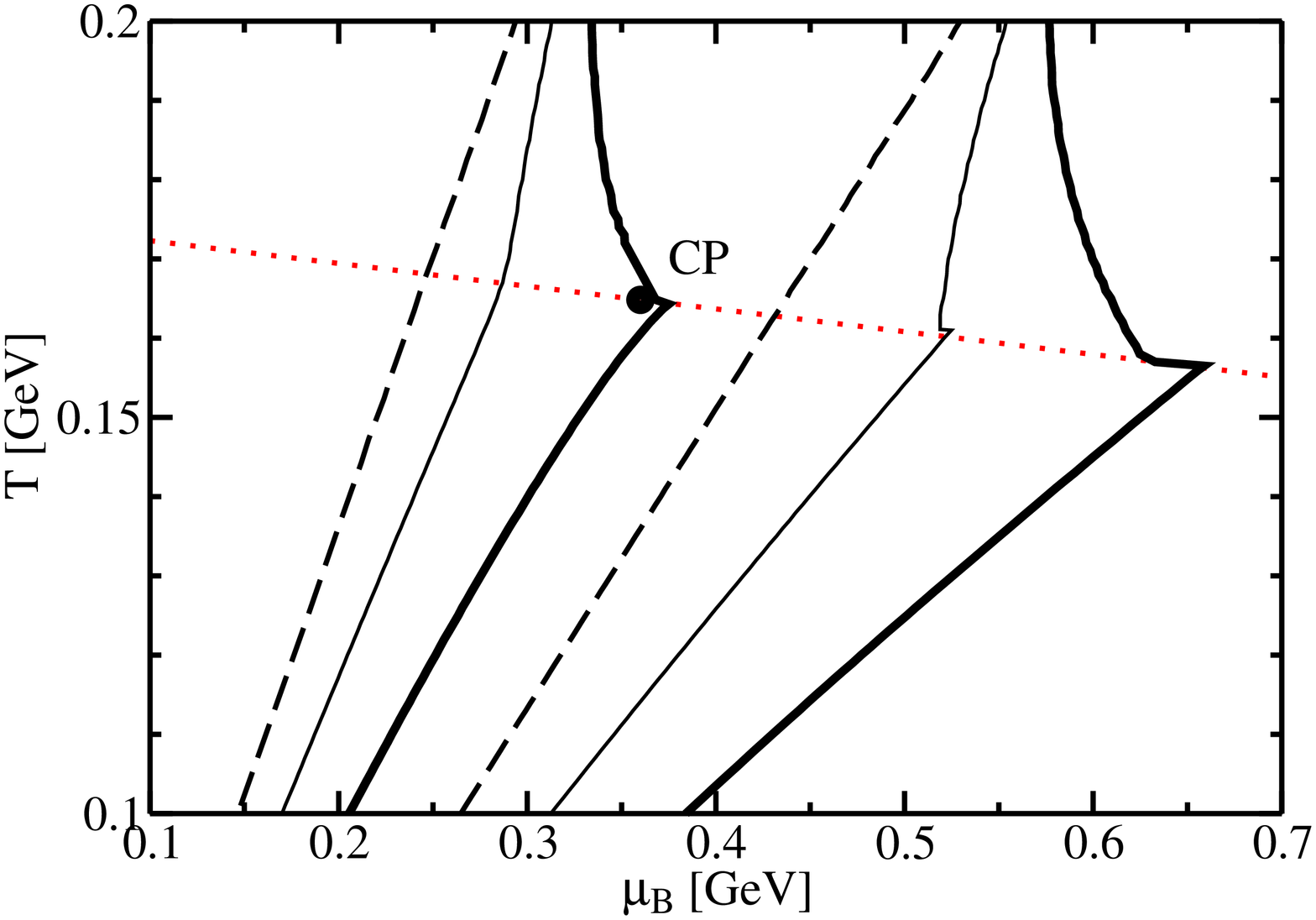}
  \caption{Influence of CP on the pattern of isentropic trajectories. Left panel: $s/n_B=300, \,45, \,35$ (from left to right) for the 
    QPM without CP (solid curves, cf. Fig.~1) and with CP (dashed curves). For small $\mu_B$ 
    the influence of CP on the EoS is negligible. CP is located according to \cite{Fodor} 
    on the estimated phase border line (dotted curve). Note that in our approach $s_{reg}$ and therefore $s_{sing}$ are based on 
    a Taylor series expansion which is limited by its radius of convergence, $\mu_B/T \le 1.8$ \cite{Alli}. It is therefore 
    impractical to examine trajectories on the right of CP within this ansatz. In contrast, the full QPM is not hampered by these 
    shortcomings \cite{Bluhm3} and it would, therefore, be conceivable to apply the full entropy density $s=\partial p/\partial T$ 
    from the QPM expression for $p$ as regular part, defining $s_{sing}=A s_{reg} tanh\,S_c$. These considerations will be 
    reported elsewhere. Here, we employ a toy model for illustration purposes in the right panel: Influence of CP on the isentropic 
    trajectories $s/n_B=50, \,28$ (from left to right) in a toy model. Here, $s_{reg}=4\bar{c_0}T^3 + \frac{2}{9}\bar{c_2}\mu_B^2 T$, 
    $s_{sing}=\frac{2}{9}\bar{c_2}\mu_B^2 T A tanh\,S_c$ with $\bar{c_0}=(32+21N_f)\pi^2 /180$, $\bar{c_2}=N_f/2$ and $N_f=2$ 
    where $A=0, 0.5, 1$ for dashed, thin and solid curves, respectively. For details cf. \cite{QM05}. Dotted curve represents 
    the $r$-axis. The isentropic trajectories display the typical pattern when crossing a first-order border line \cite{Barz} 
    signalling a finite latent heat.} 
  \label{fig:CEPisentrops}
  \end{center}
\end{figure}
trajectories \cite{QM05}. Only for larger $\mu_B$ in the vicinity of the CP, the isentropic trajectories are slightly modified, 
however, leaving the general pattern unchanged. 

Increasing $A$ or $\Delta T_c$ and $\Delta\mu_{B,c}$ and therefore the strength of 
the singular contribution or the extension of the critical region, respectively, would change the isentropic 
trajectories also for smaller $\mu_B$. For instance, in \cite{QM05} a simple toy model including 
schematically CP effects was discussed 
where for larger $A>0$ trajectories were bent into the opposite direction towards larger $\mu_B$ due to the presence of CP 
showing clear signatures of an attractor. Furthermore, trajectories on the right of CP exhibited the typical first-order 
behavior, cf. Fig.~5 (right panel). 
Note that in \cite{Nonaka05} a model was constructed involving a hadronic low-temperature and a partonic high-temperature 
phase in which CP effects also change the pattern of the trajectories. These results significantly depend on 
the entropy density, in particular in the hadronic phase, but also on the strength parameter $A$ and the extension of the 
critical region. However, lattice QCD results \cite{Ejiri} seem to exclude such strong effects of CP. 

The small effect observed for small $\mu_B$ in Fig.~5 
is understood by the numerical smallness of $s_{sing}$ (cf. Fig.~6 left panel) and its 
$\mu_B$-dependence in comparison with 
\begin{figure}[ht]
  \includegraphics[scale=0.28,angle=-90.]{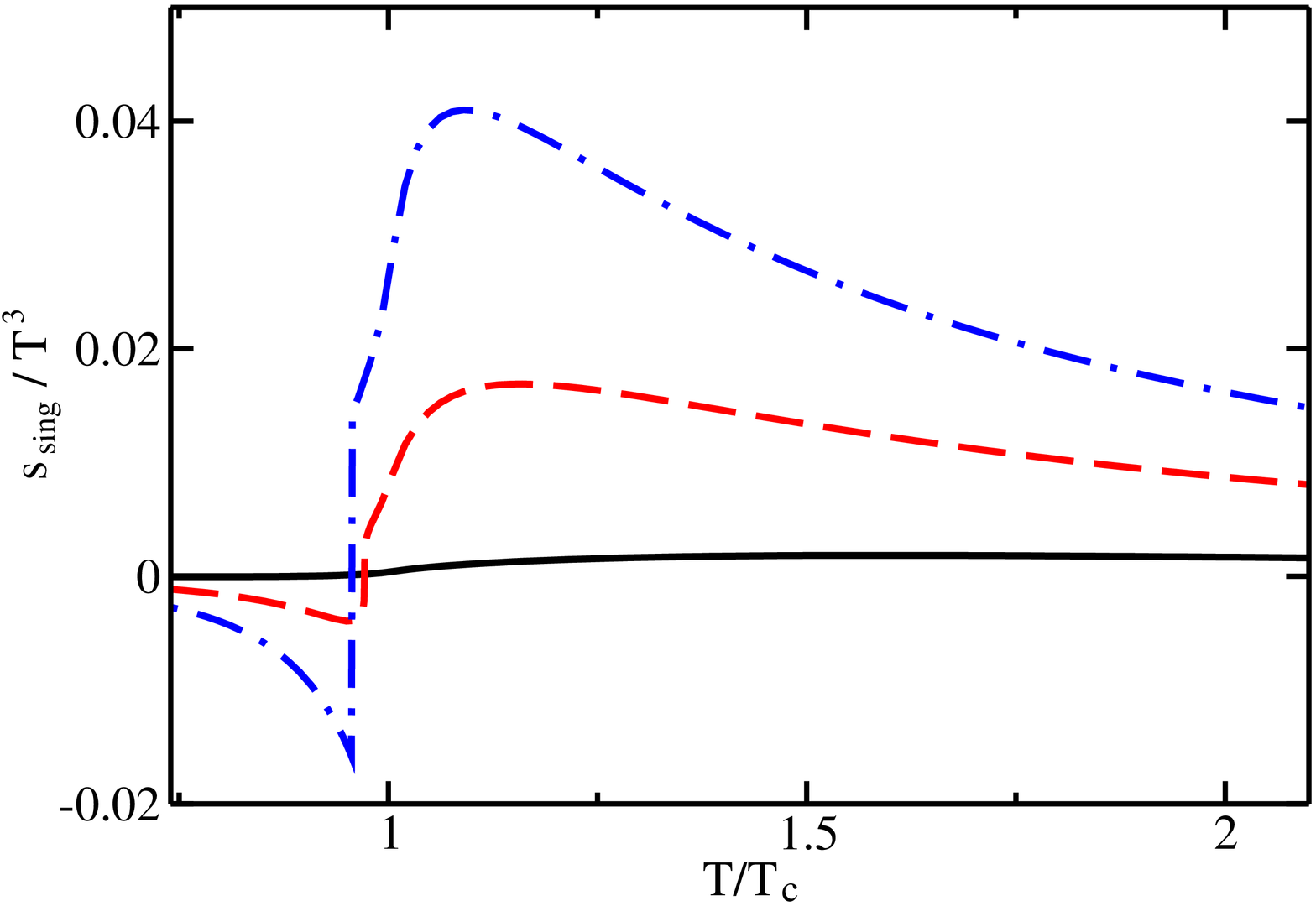}
  \includegraphics[scale=0.28,angle=-90.]{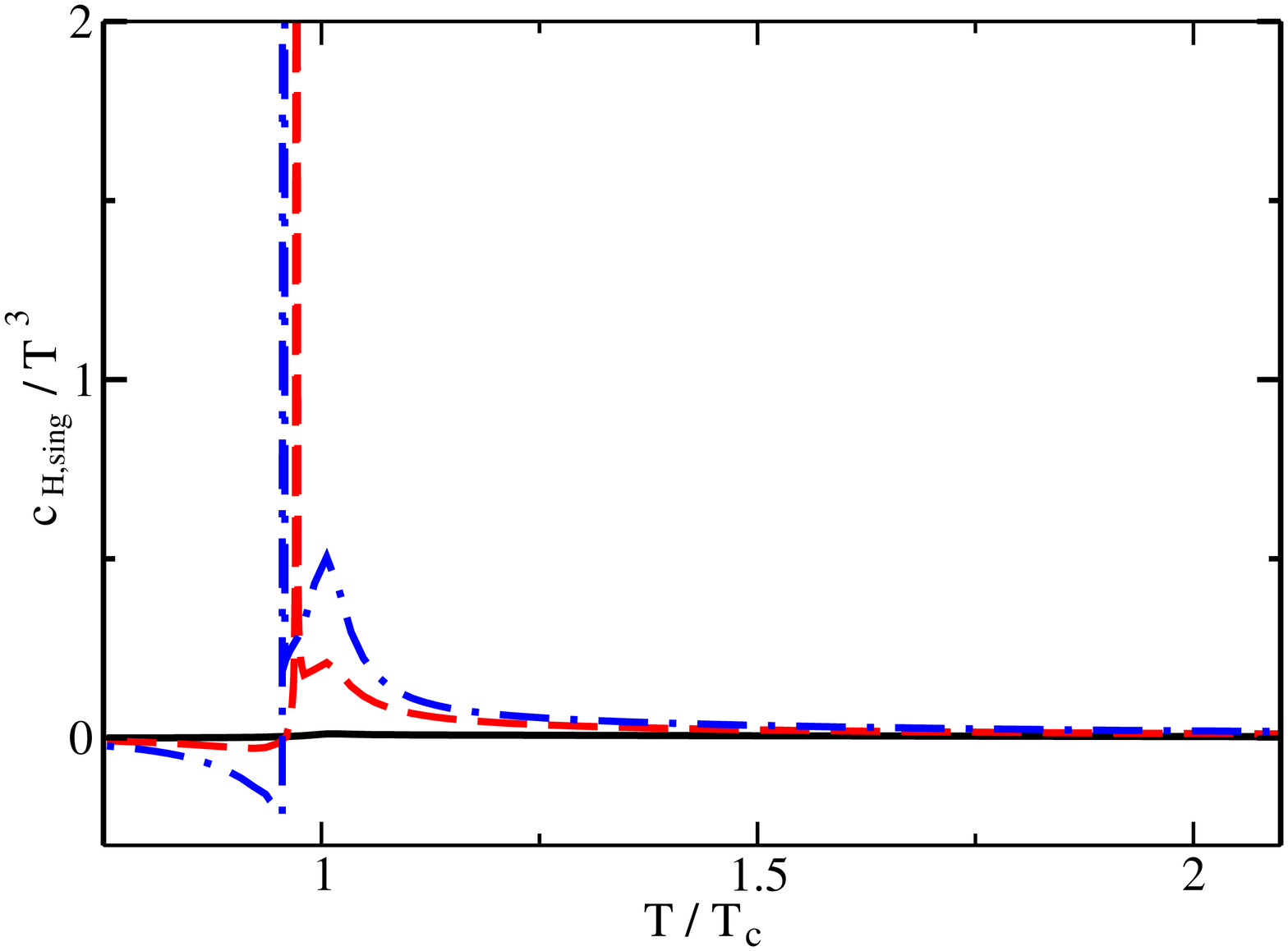}
  \caption{Left panel: scaled singular part of the entropy density $s_{sing}/T^3$ (3.7) as a function of $T/T_c$ 
    for different 
    $\mu_B=210, \,360, \,450$ MeV (solid, dashed and dash-dotted curves, respectively). For small $\mu_B$, $s_{sing}$ is 
    negligible compared to $s_{reg}$ (cf. entropy density $s(T,\mu_q=0)$ in Figure~1 of \cite{Bluhm2}). 
    At CP the slope of $s_{sing}$ becomes infinite resulting in a diverging specific heat $c_H=T\frac{\partial s}{\partial T}$. 
    For $\mu_B>360$ MeV, $s_{sing}$ shows the significant discontinuity across the phase transition line of first-order. 
    Right panel: corresponding singular part of the specific heat $c_H$ (same line code).}
  \label{fig:CEPssing}
\end{figure}
$s_{reg}$ (cf. entropy density $s(T,\mu_q=0)$ in Figure~1 of \cite{Bluhm2}). At larger values of $\mu_B$, 
the influence of CP increases, but 
with our focus on the small-$\mu_B$ region we conclude that CP effects may be small for small net baryon densities, 
in particular on averaged 
hydrodynamics. Quark-meson model based studies combined with the proper-time renormalization group method 
\cite{Scha} also point to CP effects concentrated on a fairly narrow region. 

\begin{figure}[ht]
  \includegraphics[scale=0.28,angle=-90.]{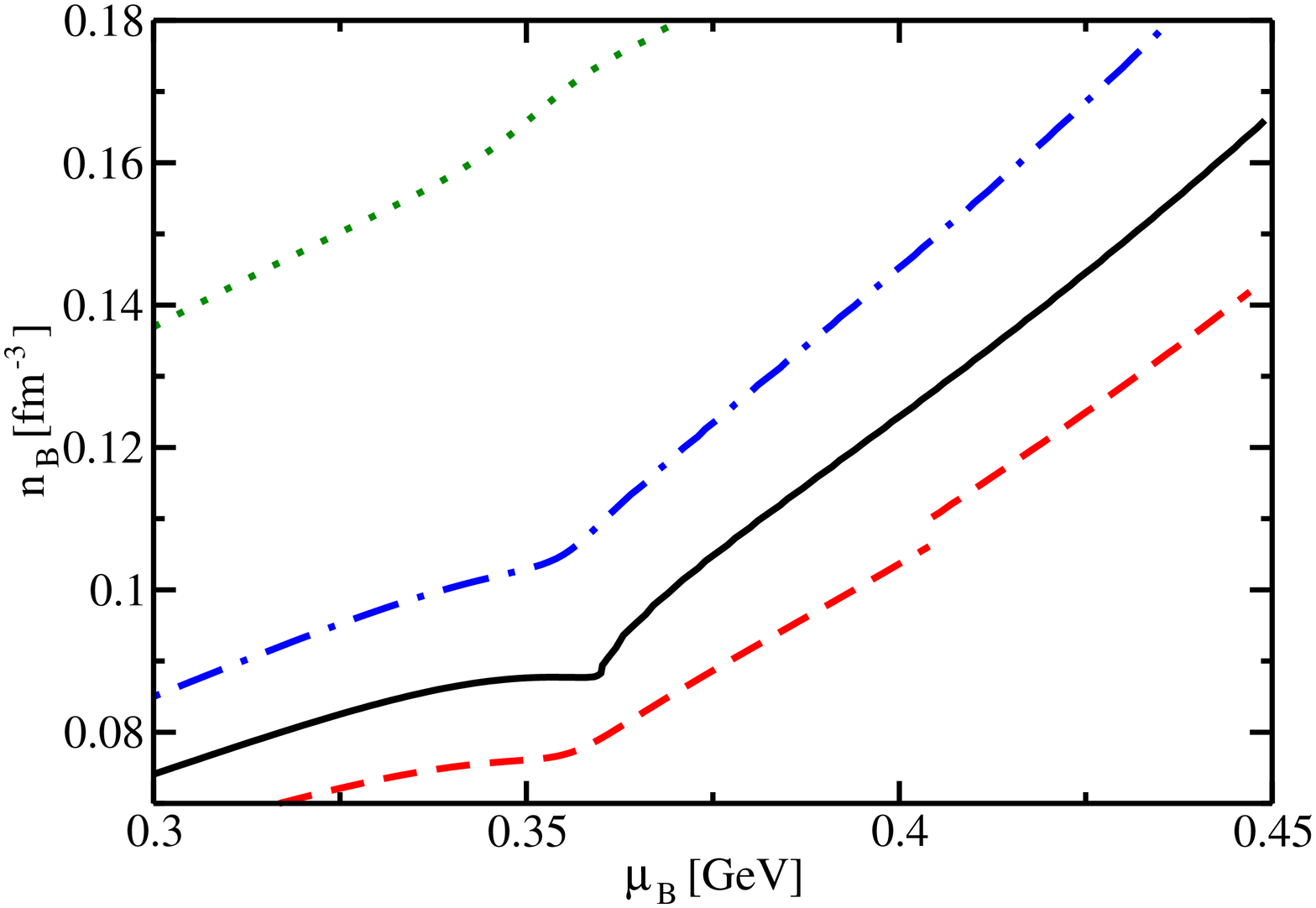}
  \includegraphics[scale=0.28,angle=-90.]{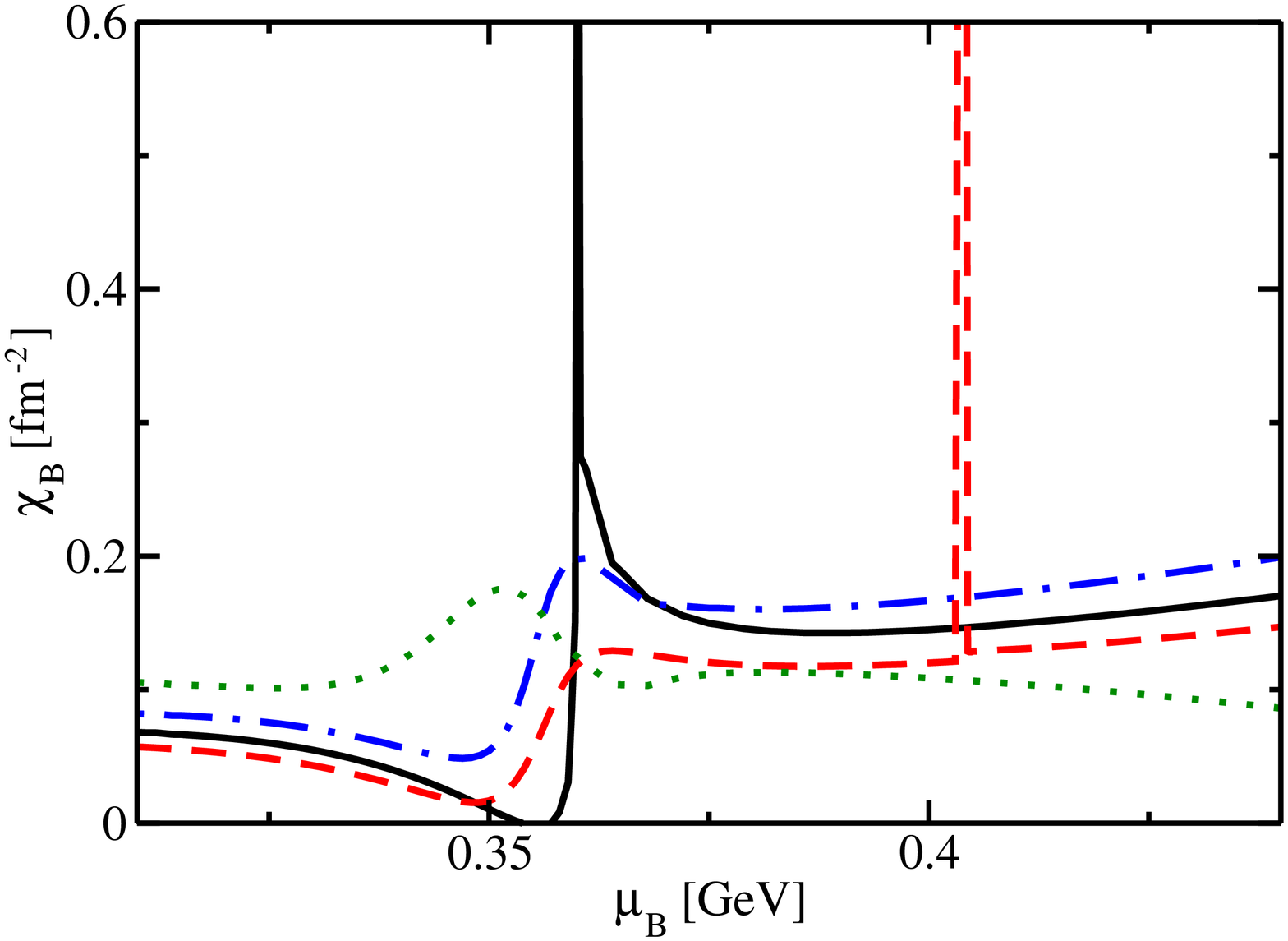}
  \caption{Left panel: baryon number density $n_B$ as a function of $\mu_B$ for different temperatures $T=T^{CP}-1.25$ MeV, $T^{CP}$, 
    $T^{CP}+1.25$ MeV and $T=176$ MeV (dashed, solid, dash-dotted and dotted curves, respectively). 
    For $T>T^{CP}$, $n_B(\mu_B)$ is continuously increasing, however showing an increasing influence of CP's 
    presence by a tiny depression when approaching $T^{CP}$; at $T=T^{CP}$, 
    the slope becomes infinite at $\mu_B^{CP +}$, 
    and for $T<T^{CP}$, $n_B(\mu_B)$ exhibits the discontinuous behavior of a first-order phase transition when 
    crossing the phase boundary. Right panel: corresponding baryon number susceptibility $\chi_B=\frac{\partial n_B}{\partial \mu_B}$ 
    as function of $\mu_B$ (same line code). The divergence of $\chi_B(\mu_B^{CP})$ is clearly seen, while the spike for 
    $T=T^{CP}-1.25$ MeV at $\mu_B = 0.41$ GeV is an artifact of the numerical differentiation at the density discontinuity. 
    Note that due to the singular contribution we are restricted in our choice of parameters in 
    (3.5, 3.7) in order to maintain thermodynamic self-consistency. The parameters here are chosen such that 
    this possibility becomes visible in the region near $\mu_B^{CP -}$, where $\chi_B(\mu_B)$ drops as a consequence 
    of the depression in $n_B(\mu_B)$. In principle, for an unfortunate parameter choice, the slope of $n_B(\mu_B)$ in the 
    vicinity of CP can become negative resulting in a negative $\chi_B$.}
  \label{fig:CEPmods}
\end{figure}

In contrast, for $\mu_B$ close to CP, baryon number susceptibility $\chi_B=(\partial n_B / \partial \mu_B)_T$ 
or specific heat $c_{H} = T (\partial s / \partial T)_{\mu_B}$ 
are strongly influenced by the QCD critical point. At CP and across the first-order phase transition line, $c_H$ diverges 
due to its singular part (cf. Fig.~6 right panel). In order to illustrate the effects on $\chi_B$, we 
increase $A=1$ which does not alter the qualitative behavior of the observables. In Fig. \ref{fig:CEPmods} (left panel), 
we exhibit the baryon number density $n_B$ as a function of $\mu_B$ which shows the expected continuous behavior in the 
crossover regime and also for $T>T_c$. 
At CP, however, the slope tremendously increases becoming infinite which turns into a finite discontinuity 
across the first-order phase transition line. Accordingly, its derivative $\chi_B$ 
as measure for fluctuations in $n_B$ diverges at CP and across the phase 
transition line whereas $\chi_B$ remains finite for temperatures $T>T^{CP}$ (cf. Fig. \ref{fig:CEPmods} right panel). 
Also, a suppression of baryon number fluctuations in the confined phase (solid, dashed and dash-dotted curves in 
Fig. \ref{fig:CEPmods} right panel) are evident in the vicinity of CP, whereas this is not the case in the 
deconfined phase (dotted curve). In contrast, the isovector susceptibility $\chi_I$ is expected to remain finite when 
approaching CP \cite{All05}. Here, we examined the behavior of $\chi_B$ for constant temperatures. In \cite{Redlich} it 
was argued that a verified signal of CP would require a non-monotonic behavior of $\chi_B$ along the phase boundary 
$T_c(\mu_B)$. 

Other possibilities of including CP in singular contributions of thermodynamic bulk quantities are conceivable, 
e.~g. $s_{sing}$ in (\ref{equ:A3}) could be chosen proportional to $\mu_B^3$ rather than $\mu_B^2 T$. Also, 
the inclusion of CP effects into different toy models were reported in \cite{QM05}. 
Different phenomenological approaches of constructing an EoS with CP behavior can be found in \cite{Brazil,Ratti}. However, 
a unique adjustment to lattice QCD data without better information or guidance seems not to be accessible. 

\section{Conclusions}

In summary we present here a phenomenologically guided construction to supplement a simple quasi-particle model 
by elements displaying the static criticality of the 3-dimensional Ising model, i.~e. the universality class QCD 
belongs to. In doing so, we apply a procedure well probed in solid state and condensed matter physics, but which is restricted 
to a narrow region around the critical point, provided its existence. Having adjusted the pure 
quasi-particle model to available lattice QCD results, little space is left for pronounced effects related to the 
critical point. Of course, an divergence of susceptibilities at the critical point and the first-order phase transition 
beyond it (at larger chemical potentials) are found, but the modification of the equation of state remains rather modest, 
as evident in the isentropic curves in the $T$ - $\mu_B$ plane. It turns out that various ans\"atze are conceivable 
and that thermodynamic consistency restricts the parameter space. The latter fact leaves us with the impression that 
a readjustment of the quasi-particle model parameters, such that even for sizeable critical point effects the lattice 
QCD results are recovered, is hardly possible. 

One major prospect of observable consequences of the critical point is the occurrence of increased fluctuations. 
An adequate theoretical study has to rely on a more microscopical framework than offered by using the equation of 
state in standard hydrodynamics, say along the lines in \cite{Brazil,Paech}. Nevertheless, the presented model may 
be used in further studies, e.~g. considering hydrodynamical expansion along neighboring paths crossing or not the 
first-order phase transition region or the critical point itself and analyzing sensible flow characteristics 
like elliptic flow and higher moments thereof. Another issue may be the study of critical point effects on penetrating 
probes, like photons and dileptons \cite{Gallmeister}, i.~e. the electromagnetic emissivity of hot and dense 
strongly interacting matter. 

\section*{Acknowledgment}

\noindent
This work is supported by BMBF 06 DR 121, 06 DR 136, GSI, Helmholtz association VI and EU-I3HP. We thank M. Asakawa, F. Karsch, 
E. Laermann, A. Peshier, K. Redlich and J. Wambach for fruitful discussions.

\end{document}